# A Novel QoE-Aware SDN-enabled, NFV-based Management Architecture for Future Multimedia Applications on 5G Systems

Alcardo Alex Barakabitze, Lingfen Sun, Is-Haka Mkwawa, Emmanuel Ifeachor
School of Computing, Electronics and Mathematics, University of Plymouth, UK
{alcardoalex.barakabitze, L.Sun, is-haka.mkwawa, E.Ifeachor}@plymouth.ac.uk

*Abstract*—This paper proposes a novel QoE-aware SDN enabled NFV architecture for controlling and managing Future Multimedia Applications on 5G systems. The aim is to improve the QoE of the delivered multimedia services through the fulfilment of personalized QoE application requirements. This novel approach provides some new features, functionalities, concepts and opportunities for overcoming the key QoE provisioning limitations in current 4G systems such as increased network management complexity and inability to adapt dynamically to changing application, network transmission/traffic or end-user's demand.

## I. INTRODUCTION

The increasing latency sensitive application scenarios such as the Internet of Things (IoT), live video streaming and virtual reality pose a challenge to service/cloud providers, and mobile network operators towards QoE provisioning to end users due to massive emerging Future Multimedia Applications (FMAs). Software Defined Networking (SDN) in 5G promises to provide and implement new capabilities and solutions for enabling network control to be programmable, centrally manageable, adaptable and cost effective which makes it suitable for bandwidth intensive applications such as video streaming. With the targeted increase of traffic diversity and capacity growth in 5G, Network Function Virtualization (NFV) will enable flexibility needed by network, cloud and mobile service providers in adapting to future dynamic market drivers. NFV in 5G will also enable network functions to be deployed dynamically to support the delivery of fast growing multimedia services.

In this paper, a novel QoE aware SDN-enabled, NFV-based management architecture for FMAs on 5G called "*QoE-Softwarized*" is proposed. The *QoE-Softwarized* should enable mobile operators and service providers to deliver quality services through an autonomic software lifecycle management approach in 5G.

## II. RELATED WORK

An SDN QoE service architecture that is suitable for LTE networks for enhancing the performance of Over-The-Top (OTT) applications has been proposed in [1]. In [2], an integrated network control and management framework for QoS support based on SDN is described. Authors in [3] proposed a management and orchestration platform for enhancing the QoS and QoE of real time multimedia applications using SDN. In [4], a holistic SDN control plane for 5G multimedia transmission engineering problem is introduced. Although some of these proposals use SDN, they are limited to QoS only. Proposals that use SDN [1, 2, 4] with QoE considerations on 5G have no or limited considerations on the integration of NFV. NFV motivates new pathways for increasing flexibility to accommodate and adapt dynamically to increasing traffic diversity. In this paper, a novel QoE aware SDN-enabled, NFV-based control and management architecture for FMAs is proposed, which aims to provide flexible networking and configurable approach for QoE provisioning through the integration of SDN and NFV.

## III. QOE-SOFTWARIZED ARCHITECTURE, CONCEPTS AND TERMINOLOGIES

Fig 1 shows the proposed *QoE Softwarized* architecture which consists of QoE management and SDN-NFV infrastructure layers enabled by SDN controllers. This approach takes the softwarization advantages by adopting the principles of SDN and NFV in order to meet the Key Performance Requirements/Indicators (KPR/I) of 5G [4]. It complies with SDN-NFVs requirements which translate also to 5G KPRs such as consistent service availability, accessibility/retainability, throughput and minimal latency.

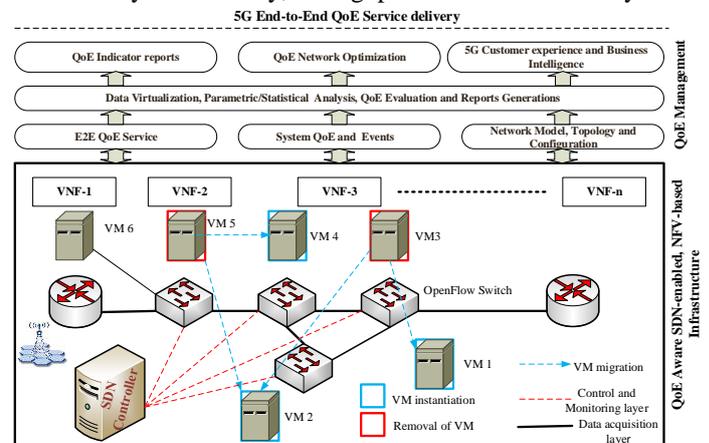

Fig 1: QoE Aware SDN enabled, NFV-based architecture for 5G systems

In the proposed architecture, the QoE-Aware SDN-NFV Infrastructure layer forms the data acquisition layer which is a set of different Virtual Network Functions (VNFs). The VNFs form a *"Quality Service Function Chain (QSFC)"* and are arranged together depending on a set of network traffic flows. SDN is responsible for the dynamic orchestration and communications between QSFC, the network apps/services in the cloud and user's mobile terminal. The *QoE-Softwarized* architecture consists of VMs which are deployed in any location within the SDN-NFV infrastructure such that each VM will obtain different requirements for network resources allocation and utilization. The VMs should have the required quality measures/parameters of QoS and QoE of a particular multimedia application specifically in terms of network bandwidth, scheduling latency, jitter and stalling etc. Based on the decoupling of services from resources offered by NFV and SDN paradigm by detaching lifecycle management from physical constraints, the locations of VNFs may change in the network. For example, during malfunctioning of any VM, the affected VNFs can be migrated to another location in order to avoid service delivery interruptions. In such a case, SDN controller is used to allocate paths and provide connectivity dynamically among VNFs and therefore acting as an enabler of NFVs. SDN provides an ability to realize the multimedia network services by dynamically chaining together VNFs and direct multimedia service traffic flows into those VNFs chains. In order to enable interactions between VNFs and SDN controller for the overall quality service function chain, a QoE-aware SDN controller structure as shown in Fig.2 is proposed to enable the delivery of multimedia services over 5G networks with optimized or acceptable QoE.

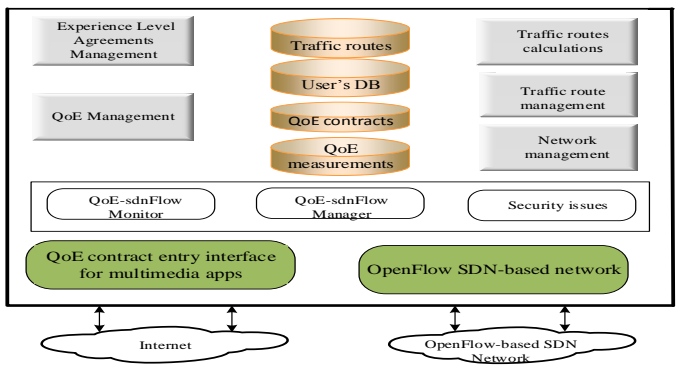

Fig 2: SDN Controller with QoE Entities

The QoE control and management of multimedia services in different network domains will be made through a well-defined Service Level Agreement (SLA) or using the more recent QoE-oriented Experience Level Agreements (ELAs) [5]. The *QoE-sdnFlow Monitor* and QoE–sdnFlow manager perform the QoE estimation and measurements per multimedia traffic flow. They acquire network topology information and implement QoE based network policies by using different control algorithms for traffic prediction, admission control, radio resource allocation, load balancing and user density prediction. In addition to QoE related blocks, other functionalities such as security will also be taken into account in the SDN controller.

### IV. APPLICATION SCENARIOS AND USE CASES

The proposed *QoE-Softwarized* architecture may be used in a wide range of applications. Fig 3 shows an application scenario where a user requests a multimedia application from a service provider.

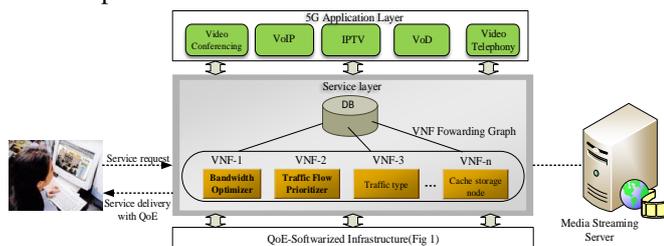

Fig 3. Multimedia Application User's Request Scenario

The service layer consists of a range of VNFs. Depending on the type of multimedia application (e.g., IPTV, VoIP, live video conferencing) requested by the user, the network service functions provide a set of functional operations to the user and fulfil his/her personalized QoE application requirements. In this scenario, we consider a chain of VNFs that support bandwidth and high quality demanding applications such as live video conferencing and IPTV in home networks. For the case of live conferencing, the *QoE-Softwarized* should be able to provide sufficient QoE through multi-point to multi-point connectivity among conference attendees using a set of VNFs. The user's request is transported across these VNFs. Depending on the user's request, the VNFs then formulates a service graph which establishes the VNFs Forwarding Graph stored in the VNFs database (DB). The QoE-sdnFlow Manager or the SDN-NFV orchestrator instantiates the DB as soon as the user requests a multimedia application with specific QoS and QoE indices. It will react accordingly based on feedback obtained from the QoE-sdnFlow Monitor in order to maintain/optimise a user's end-to-end QoE.

### V. CONCLUSIONS

This paper introduced the concept of *QoE-Softwarization* and proposed the QoE-aware SDN-NFV architecture for delivering multimedia services over 5G networks. New QoE related entities, such as QoE-sdnFlow monitor and manager are introduced to implement QoE related policies and techniques in an SDN controller. The proposed *QoE-Softwarized* architecture would help in designing and implementing QoE control and management schemes for future multimedia services.

### ACKNOWLEDGMENT

The work presented in this paper is funded by the European Union in the context of Horizon2020 Research and Innovation Programme under Marie Skłodowska-Curie Innovative Training Networks (MSCA-ITN-2014-ETN), Grant Agreement No.643072, Network QoE-NET.